\documentclass[showkeys,preprintnumbers,amsmath,amssymb]{revtex4-2}

\usepackage[utf8]{inputenc}
\usepackage[pdftex]{graphicx}
\usepackage{caption}
\usepackage{subcaption}
\usepackage{mathrsfs}
\usepackage[colorlinks, breaklinks, urlcolor={blue}, linkcolor={red}, citecolor={blue}]{hyperref}
\usepackage{array}
\usepackage{amsmath}
\usepackage{type1cm}
\usepackage[export]{adjustbox}
\usepackage{dsfont}
\usepackage{lettrine}
\usepackage[english]{babel}
\usepackage{lmodern}
\usepackage{microtype}
\usepackage{booktabs}
\usepackage[T1]{fontenc}
\usepackage[boxed, vlined]{algorithm2e}
\usepackage{braket}
\usepackage{xcolor}
\usepackage{orcidlink}
\usepackage{braket}
\usepackage{bm}
\usepackage{bbold}
\usepackage{tikz}
\usepackage[utf8]{inputenc}

\usepackage{upgreek}
\usepackage{xcolor}
\usepackage{soul}

\begin{document}
	
	\title{Robustness of Quantum Discord in Nonequilibrium Electronic Transport through Tunnel-Coupled Quantum Dots}
	
	\author{Thingujam Yaiphalemba Meitei}
	\affiliation{Department of Physics and Nanotechnology, SRM Institute of Science and Technology, Kattankulathur 603203, Tamil Nadu, India.}
	
	\author{Saikumar Krithivasan}
	\affiliation{Department of Physics and Nanotechnology, SRM Institute of Science and Technology, Kattankulathur 603203, Tamil Nadu, India.}
	
	\author{Arijit Sen\orcidlink{0000-0002-8624-9418}}
	\email[]{arijits@srmist.edu.in}
	\affiliation{Department of Physics and Nanotechnology, SRM Institute of Science and Technology, Kattankulathur 603203, Tamil Nadu, India.}
	
	\author{Md~Manirul Ali  \orcidlink{0000-0002-5076-7619}}
	\email[]{manirul@citchennai.net}
	\affiliation{ Centre for Quantum Science and Technology, Chennai Institute of Technology, Chennai 600069, India.
	}
	
	\begin{abstract}
		Quantum discord captures quantum correlations beyond entanglement and can remain finite even when the entanglement vanishes. We investigate the transient nonequilibrium dynamics and steady-state behavior of quantum discord and classical correlations in a double quantum dot (DQD) system coupled to fermionic reservoirs. By employing a quantum Langevin equation formalism, we obtain the exact reduced density matrix of the system, enabling a comprehensive analysis of its quantum and classical correlations under nonequilibrium conditions. The influence of system-reservoir coupling strength, spectral bandwidth, thermal bias, and varying initial state on both the transient dynamics and steady-state correlations is systematically analyzed. Quantum discord remains finite in the nonequilibrium steady state over a broad parameter range. Although thermal gradients reduce the overall magnitude of correlations, quantum discord persists and exhibits greater resilience. These results demonstrate that nonequilibrium electronic transport, together with the environmental spectral properties and reservoir asymmetry, provides an effective means of controlling nonclassical correlations in mesoscopic systems and establishes quantum discord as a robust hallmark of open fermionic quantum devices.
	\end{abstract}
	
	\maketitle
	
	\section{Introduction}
	
	Understanding the behavior of nonclassical correlations in nonequilibrium open quantum systems is a central problem in condensed matter physics~\cite{breuer2002theory, schaller2014open}, and is closely related to their role in quantum information processing and emerging quantum technologies~\cite{ali2026near, NielsenChuang2000, Horodecki2009}. Mesoscopic and nanoscale systems coupled to external reservoirs form an ideal platform to study the interplay between quantum transport and quantum resources. While entanglement has long been regarded as the hallmark of quantum correlations, it is now well established that more general forms of quantum correlations exist and can remain finite even in regimes where entanglement vanishes~\cite{adesso2016measures, elghaayda2022local, cao2020fragility}. These correlations, often quantified through measures such as quantum discord, have attracted significant attention in dissipative and nonequilibrium quantum systems.
	
	Quantum discord, introduced independently by Ollivier and Zurek~\cite{OllivierZurek2001} and by Henderson and Vedral~\cite{ HendersonVedral2001}, quantifies quantum correlations beyond entanglement as the difference between total correlations and those that can be extracted through local measurements. Importantly, it can remain nonzero even in separable states, thereby revealing correlations inaccessible to entanglement measures. This feature has motivated extensive studies of quantum discord in diverse contexts, such as quantum computation~\cite{datta2008quantum, datta2011quantum}, metrology~\cite{sone2019nonclassical, singh2014quantum, hunt2019observe}, communication~\cite{madhok2013quantum}, cryptography~\cite{pirandola2014quantum}, and thermodynamics~\cite{jimenez2019quantum}. More recent works have demonstrated its robustness in optomechanical systems~\cite{chabar2025gaussian} and strongly correlated open systems~\cite{jafari2025dynamics}, established through analytical bounds and detection schemes~\cite{kibret2024generation}. Experimental studies across a wide range of platforms further confirm its relevance in realistic quantum systems~\cite{xu2010experimental, auccaise2011experimentally, madhok2018quantum, pirandola2014optimality, gulati2025experimental}.
	
	Fermionic systems provide a particularly relevant setting for investigating quantum correlations under realistic nonequilibrium conditions~\cite{faba2021two, wang2019nonequilibrium}. Owing to their intrinsic anticommutation relations, fermionic systems exhibit distinct statistical and structural features that fundamentally influence correlation properties~\cite{Friis2010, Benatti2004, cialdi2019experimental}. Previous studies have explored quantum discord in relativistic and non-inertial systems~\cite{Datta2008, WangJing2011, Hu2012}, spin models~\cite{huang2014scaling, yurishchev2011quantum, dillenschneider2008quantum}, and open quantum systems under environmental noise~\cite{Maziero2009}, consistently demonstrating its persistence in regimes where entanglement is suppressed.
	
	Solid-state platforms based on double quantum dot (DQD) systems provide a controllable environment for investigating quantum correlations at the mesoscopic scale~\cite{LossDiVincenzo1998, Hanson2007, Koppens2006, meitei2024quantumness}. These systems allow precise tuning of tunneling, interactions, and coupling to electronic reservoirs, enabling systematic studies of the interplay between transport~\cite{Petta2005, Burkard1999, sen2010effect}, decoherence~\cite{chaouki2022dynamics, ali2022quantum}, and quantum correlations~\cite{wang2025quantum, ait2023global, ait2024prospecting}. In particular, tunnel-coupled quantum dots connected to fermionic reservoirs constitute a natural platform for exploring nonequilibrium dynamics driven by system-environment interactions, thermal gradients, and reservoir memory effects. Previous works have primarily focused on entanglement and coherence in such systems under equilibrium and nonequilibrium conditions~\cite{Werlang2009, Fanchini2010, Maziero2009, Xu2010, RoszakMachnikowski2006, Bellomo2007, miao2025entanglement, sadeghi2025quantum}. However, it has become increasingly clear that quantum correlations beyond entanglement can exhibit enhanced resilience to dissipative processes~\cite{Ferraro2010, LoFranco2012}.
	
	Despite these advances, a systematic understanding of quantum discord in nonequilibrium tunnel-coupled quantum dots, particularly in the presence of structured fermionic reservoirs, remains incomplete. In particular, it is not yet clear how reservoir memory effects determine the dynamics and steady-state behavior of quantum correlations. In this work, we investigate the nonequilibrium dynamics of quantum discord in tunnel-coupled quantum dots connected to independent fermionic reservoirs, which serve as a prototypical platform for mesoscopic transport. Using an exact reduced density matrix obtained through the quantum Langevin equation with Lorentzian spectral densities, we analyze the effects of system-reservoir coupling strength, spectral bandwidth, and temperature gradients on both transient dynamics and steady-state correlations.
	
	We identify a nonequilibrium steady-state regime in which quantum discord remains finite, establishing a distinct regime of quantum correlations in a mesoscopic fermionic system. The persistence of these correlations is governed by the interplay of environmental spectral properties, reservoir memory, and coupling asymmetry, providing a route for controlling quantum correlations in solid-state devices. This behavior goes beyond the conventional robustness of quantum discord under dissipation and highlights the role of nonequilibrium environments in stabilizing quantum correlations in open fermionic transport systems.
	
	\section{Model and Theoretical Framework}
	
	We employ an exact reduced density matrix that fully captures system-reservoir interactions while retaining non-Markovian memory effects without invoking Markovian approximations.
	\subsection{Nonequilibrium Dynamics of Tunnel-Coupled Quantum Dots}
	We consider a double quantum dot connected to two fermionic reservoirs maintained at different chemical potentials.  
	The total Hamiltonian, $H$, is given by
	\begin{equation}
		H = H_S + H_B + H_I,
	\end{equation}
	where $H_S$ is the Hamiltonian of the double-dot structure: $H_S = \sum_{i,j=1}^{2} \epsilon_{ij} a_j^\dagger a_i$. The on-site energies of the first and second quantum dots are represented by $\epsilon_{11}$ and $\epsilon_{22}$ respectively, while the cross-terms $\epsilon_{12}$ and $\epsilon_{21}$ represent hopping energies that allow a particle to tunnel between dots. The fermionic creation and annihilation operators $a_j^\dagger$ and $a_j$ correspond to the $j^{\mathrm{th}}$ dot. The bath Hamiltonian $H_B = \sum_{\alpha, k} \epsilon_{\alpha k} c_{\alpha k}^\dagger c_{\alpha k}$, where $\alpha \in \{L, R\}$ labels the left or right reservoir, $\epsilon_{\alpha k}$ is the $k^{\mathrm{th}}$ energy level in reservoir $\alpha$, and $c_{\alpha k}^\dagger$, $c_{\alpha k}$ are the corresponding fermionic operators. The interaction Hamiltonian $H_I = \sum_{i,\alpha,k} (V_{i\alpha k} a_i^\dagger c_{\alpha k} + V_{i\alpha k}^* c_{\alpha k}^\dagger a_i)$ describes the tunneling-mediated exchange of particles between the double quantum dot and the reservoirs, with $V_{i\alpha k}$ denoting the coupling between the $i^{\mathrm{th}}$ dot and the $k^{\mathrm{th}}$ level of reservoir $\alpha$.
	
	The evolution of the system operators in the Heisenberg picture is governed by coupled differential equations~\cite{meitei2024quantumness}. By solving the reservoir operator equation and substituting into the system's equation of motion, we obtain the quantum Langevin equation as
	\begin{align}
		\frac{d}{dt} a_i(t) &= -i \sum_j \epsilon_{ij} a_j(t)
		- \sum_{\alpha,j} \int_{t_0}^t d\tau\, g_{\alpha ij}(t,\tau) a_j(\tau)
		\nonumber \\
		&\quad - i \sum_{\alpha k} V_{i\alpha k}\, c_{\alpha k}(t_0)
		e^{-i\epsilon_{\alpha k}(t-t_0)},
		\label{eq:langevin}
	\end{align}
	where the first term describes coherent inter-dot tunneling, while the second and third terms encode dissipation and fluctuation effects induced by the reservoirs. The memory kernel $g_{\alpha ij}(t,\tau)$ captures non-Markovian effects and is defined via the spectral density $J_{\alpha ij}(\epsilon)$ as
	\begin{equation}
		g_{\alpha ij}(t,\tau) = \int \frac{d\epsilon}{2\pi}\,
		J_{\alpha ij}(\epsilon)\, e^{-i\epsilon(t-\tau)}.
	\end{equation}
	The Langevin equation is linear and admits the formal solution as
	\begin{equation}
		a_i(t) = \sum_j u_{ij}(t,t_0)\, a_j(t_0) + F_i(t),
	\end{equation}
	where $u_{ij}(t,t_0)$ is the retarded Green's function and $F_i(t)$ is the noise operator. Assuming the system and reservoirs are initially isolated, with reservoirs in thermal equilibrium, the noise operator is given by
	\begin{equation}
		F_i(t) = -i \sum_{j,\alpha,k} \int_{t_0}^{t} d\tau\,
		u_{ij}(t,\tau)\, V_{j\alpha k}\, c_{\alpha k}(t_0)\,
		e^{-i\epsilon_{\alpha k}(\tau-t_0)}.
	\end{equation}
	The noise correlation functions describe the impact of the reservoir-induced fluctuations on the system in the following way
	\begin{eqnarray}
		\label{vij}
		&&\langle F_{j}^\dagger(t_2) F_{i} (t_1) \rangle = v_{ij}(t_1,t_2) \nonumber \\
		&&= \sum_{\alpha} \int_{t_0}^{t_1} d\tau_1 \int_{t_0}^{t_2} d\tau_2 \Big[ {\bf u}(t_1,\tau_1)
		{\widetilde{\bf g}}_{\alpha} (\tau_1,\tau_2) {\bf u}^{\dag}(t_2,\tau_2) \Big]_{ij}, \nonumber \\
	\end{eqnarray}
	and
	\begin{eqnarray}
		\label{vijbar}
		&&\langle F_{i} (t_1) F_{j}^\dagger(t_2)  \rangle = {\overline v}_{ij}(t_1,t_2) \nonumber \\
		&&= \sum_{\alpha} \int_{t_0}^{t_1} d\tau_1 \int_{t_0}^{t_2} d\tau_2 \Big[ {\bf u}(t_1,\tau_1) {\overline{\bf g}}_{\alpha} (\tau_1,\tau_2) {\bf u}^{\dag}(t_2,\tau_2) \Big]_{ij} \nonumber, \\
	\end{eqnarray}
	where the time correlation functions are given by
	\begin{eqnarray}
		\label{gtilde}
		\!\!\!\!\!\!{\widetilde{g}}_{\alpha m n} (\tau_1,\tau_2) \!=\! \sum_k V_{m \alpha k } V_{n \alpha k }^\ast f_{\alpha}(\epsilon_{\alpha k})
		e^{-i \epsilon_{\alpha k} (\tau_1-\tau_2)},
	\end{eqnarray}
	
	\begin{eqnarray}
		\label{gtbar}
		\!\!\!\!\!\!{\overline{g}}_{\alpha m n}\!(\tau_1,\tau_2) \!=\!\!\! \sum_k \!V_{m \alpha k } V_{n \alpha k }^\ast (1\!\!-\!\!f_{\alpha}(\epsilon_{\alpha k})) e^{-i \epsilon_{\alpha k} (\tau_1-\tau_2)}.
	\end{eqnarray}
	In the continuum limit these become
	\begin{eqnarray}
		\widetilde{\mathbf{g}}_\alpha(\tau_1,\tau_2) &=& \int
		\frac{d\epsilon}{2\pi}\, \mathbf{J}_\alpha(\epsilon)\, f_\alpha(\epsilon)\,
		e^{-i\epsilon(\tau_1-\tau_2)},
		\label{gtilde2}
	\end{eqnarray}
	\begin{eqnarray}
		\bar{\mathbf{g}}_\alpha(\tau_1,\tau_2) &=& \int
		\frac{d\epsilon}{2\pi}\, \mathbf{J}_\alpha(\epsilon)\,
		[1 - f_\alpha(\epsilon)]\, e^{-i\epsilon(\tau_1-\tau_2)},
		\label{gtbar2}
	\end{eqnarray}
	where $f_\alpha(\epsilon) = 1/[e^{(\epsilon-\mu_\alpha)/kT_\alpha}+1]$ is the Fermi--Dirac distribution at temperature $T_\alpha$ and chemical potential $\mu_\alpha$.
	
	For the fermionic reservoirs the spectral density is chosen as Lorentzian~\cite{Fanchini2010, LoFranco2012, krithivasan2025interplay}
	\begin{equation}
		J_{\alpha ij}(\epsilon) = \frac{\Gamma^\alpha_{ij} W_\alpha^2}
		{(\epsilon - \mu_\alpha)^2 + W_\alpha^2},
	\end{equation}
	with bandwidth $W_\alpha$ and coupling strengths $\Gamma^\alpha_{ij}$.
	For quantum dots coupled in series we take
	$\Gamma^L_{11} = \Gamma^L_{22} = \Gamma_L$,
	$\Gamma^R_{11} = \Gamma^R_{22} = \Gamma_R$, and
	$\Gamma^L_{12} = \Gamma^L_{21} = \Gamma^R_{12} = \Gamma^R_{21} = 0$.
	
	We describe the DQD system using the fermionic occupation-number basis. The four basis states are $|11\rangle$, $|10\rangle$, $|01\rangle$, and $|00\rangle$, representing both dots occupied, only the first occupied, only the second occupied, and both unoccupied, respectively. The fermionic operators $a_i^\dagger$, $a_i$ satisfy the canonical anticommutation relations $\{a_i, a_j^\dagger\} = \delta_{ij}$, $\{a_i, a_j\} = 0$, and $\{a_i^\dagger, a_j^\dagger\} = 0$~\cite{Friis2010, Benatti2004}.
	
	The density operator $\rho(t)$ of the DQD is a $4\times4$ Hermitian matrix whose elements are constructed from expectation values of bilinear combinations of fermionic operators in the Heisenberg picture~\cite{krithivasan2025interplay}. The diagonal elements correspond to the occupation probabilities of the four configurations so that
	\begin{align}
		\rho_{11}(t) &= \langle N_1 N_2 \rangle_t, \\
		\rho_{22}(t) &= \langle N_1(1 - N_2) \rangle_t, \\
		\rho_{33}(t) &= \langle (1 - N_1) N_2 \rangle_t, \\
		\rho_{44}(t) &= \langle (1 - N_1)(1 - N_2) \rangle_t,
	\end{align}
	where $N_i = a_i^\dagger a_i$ is the occupation operator of the $i^{\mathrm{th}}$ quantum dot.
	Substituting the Heisenberg-picture expressions for $a_i^\dagger$ and $a_i$ into $\rho_{ij}(t)$ yields both unitary contributions (through $u_{ij}$) and non-unitary, noise-induced terms (through $F_i$). The explicit expression for $\rho_{11}(t)$ is given by
	\begin{eqnarray}
		\rho_{11}(t) &=&
		u_{12}^*(t,t_0) u_{11}(t,t_0) u_{21}^*(t,t_0) u_{22}(t,t_0) \nonumber \\
		&+& |u_{12}(t,t_0)|^2 |u_{22}(t,t_0)|^2 \nonumber \\
		&+& |u_{12}(t,t_0)|^2 v_{22}(t) \nonumber \\
		&+& u_{11}(t,t_0) u_{21}^*(t,t_0) v_{21}(t) \nonumber \\
		&+& u_{12}^*(t,t_0) u_{22}(t,t_0) \bar{v}_{12}(t) \nonumber \\
		&+& |u_{22}(t,t_0)|^2 v_{12}(t) \nonumber \\
		&+& \langle F_1^\dagger F_1 F_2^\dagger F_2 \rangle_t.
	\end{eqnarray}
	The corresponding expressions for $\rho_{22}$, $\rho_{33}$, and $\rho_{44}$ follow analogously and are provided in Ref.~\cite{krithivasan2025interplay}. These elements represent the joint occupation probabilities of the two quantum dots, with $\rho_{11}$ corresponding to double occupation, $\rho_{22}$ and $\rho_{33}$ to single occupation, and $\rho_{44}$ to the empty configuration. The off-diagonal elements encode coherences between different dot configurations. Among the off-diagonal elements, only $\rho_{23}(t)$ and $\rho_{32}(t)=\rho_{23}^*(t)$ are nonzero. The former is given by
	\begin{equation}
		\rho_{23}(t) = \langle a_2^\dagger a_1 \rangle_t
		= u_{12}(t,t_0) u_{22}^*(t,t_0) + v_{12}(t).
	\end{equation}
	
	The reduced density matrix at time $t$ therefore takes the form
	\begin{equation}
		\rho(t) =
		\begin{bmatrix}
			\rho_{11}(t) & 0            & 0            & 0 \\
			0            & \rho_{22}(t) & \rho_{23}(t) & 0 \\
			0            & \rho_{32}(t) & \rho_{33}(t) & 0 \\
			0            & 0            & 0            & \rho_{44}(t)
		\end{bmatrix}.
	\end{equation}
	All dynamical quantities are evaluated numerically from the time-dependent Green’s functions and noise correlations defined above.

	\subsection{Quantum Discord and Correlation Measures}
	Quantum discord provides an entropic measure of quantum correlations beyond entanglement~\cite{OllivierZurek2001,HendersonVedral2001}. For a bipartite state $\rho_{AB}$ (where subsystems $A$ and $B$ correspond to the first and second quantum dot, respectively), the total correlation is quantified by the quantum mutual information:
	\begin{equation}
		\mathcal{I}(\rho_{AB}) = S(\rho_A) + S(\rho_B) - S(\rho_{AB}),
		\label{eq:mutualinfo}
	\end{equation}
	where $S(\rho) = -\mathrm{Tr}[\rho \log_2 \rho]$ is the von Neumann entropy, and $\rho_A = \mathrm{Tr}_B(\rho_{AB})$, $\rho_B = \mathrm{Tr}_A(\rho_{AB})$ are the reduced density matrices of subsystems $A$ and $B$.
	
	The classical correlation is obtained via projective measurements on either of the subsystem. In our study, we choose projective measurement $\{\Pi_k^A\}$ to act on subsystem $A$. The corresponding conditional entropy is $S(\rho_{B|A}) = \sum_k p_k S(\rho_A^k)$, where
	\begin{align}
		p_k = \mathrm{Tr}\!\left[
		\left(\Pi^A_k \otimes \mathbb{I}_B\right)\rho_{AB}
		\left(\Pi^A_k \otimes \mathbb{I}_B\right)
		\right], \\
		\rho^k_B = \frac{
			\mathrm{Tr}_A\!\left[
			\left(\Pi^A_k \otimes \mathbb{I}_B\right)\rho_{AB}
			\left(\Pi^A_k \otimes \mathbb{I}_B\right)
			\right]
		}{p_k}.
	\end{align}
	The classical correlation is defined as the maximum information about subsystem $B$ accessible through local measurements on $A$:
	\begin{equation}
		C(\rho_{AB}) = S(\rho_B) - \min_{\{\Pi_k^A\}} S(\rho_{B|A}).
		\label{eq:classical}
	\end{equation}
	For the structure of $\rho(t)$ in Eq.~(19), the minimization over projective measurements in Eq.~(\ref{eq:classical}) can be carried out analytically (Appendix \ref{app:discord}). Quantum discord is then defined as the difference between total and classical correlations:
	\begin{equation}
		\mathcal{D}(\rho_{AB}) = \mathcal{I}(\rho_{AB}) - C(\rho_{AB}).
		\label{eq:discord}
	\end{equation}
	The time-dependent reduced density matrix thus enables direct evaluation of both quantum discord and classical correlations, providing a unified framework for analyzing their nonequilibrium dynamics across varying coupling strengths, spectral bandwidths, and thermal biases.
	
	\section{Results and Discussion}

	\begin{figure}[h]
		\centering
		\includegraphics[width=0.9\textwidth]{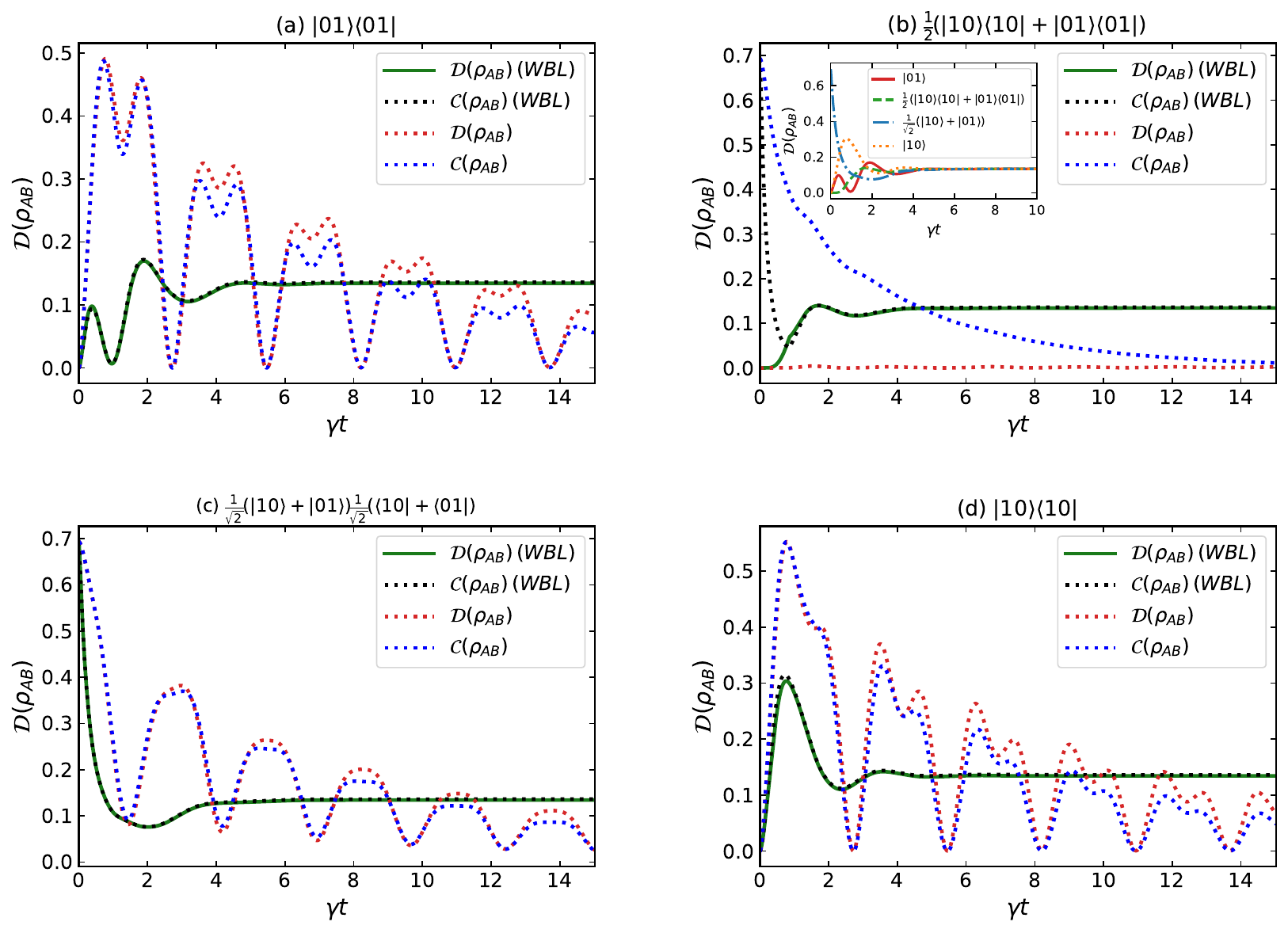}
		\caption{Time evolution of quantum discord $\mathcal{D}(\rho_{AB})$ and classical correlation $\mathcal{C}(\rho_{AB})$ for the full non-Markovian dynamics (red and black dotted, $\Gamma = \gamma$, $W = \gamma$) and the wideband limit (WBL, green solid, $\Gamma = \gamma$, $W = 50\gamma$) for four initial states: (a) $|01\rangle\langle01|$, (b) $\frac{1}{2}(|10\rangle\langle10| + |01\rangle\langle01|)$, (c) $\frac{1}{\sqrt{2}}(|10\rangle + |01\rangle)\frac{1}{\sqrt{2}}(\langle10| + \langle01|)$, and (d) $|10\rangle\langle10|$.}
		\label{fig1}
	\end{figure}

	Fig.~\ref{fig1} compares the time evolution of quantum	discord $\mathcal{D}(\rho_{AB})$ and classical correlation	$\mathcal{C}(\rho_{AB})$ in the full non-Markovian dynamics	($\Gamma = \gamma$, $W = \gamma$) and the wideband limit
	(WBL, $\Gamma = \gamma$, $W = 50\gamma$) across four	initial states: the product states $|01\rangle\langle01|$
	and $|10\rangle\langle10|$, the superposition state $\frac{1}{\sqrt{2}}(|10\rangle + |01\rangle) \frac{1}{\sqrt{2}}(\langle10| + \langle01|)$, and the
	classical mixture	$\frac{1}{2}(|10\rangle\langle10| + |01\rangle\langle01|)$.
	The four initial states are chosen to systematically study	the role of initial states in	shaping the transient and steady-state behavior of quantum	discord and classical correlation. All energies are expressed in units of $\gamma$, which sets the natural energy scale. Unless stated otherwise, all calculations are performed with dot energy levels $\epsilon_{11} = 3\gamma$ and $\epsilon_{22} = 2\gamma$, interdot coupling	$\epsilon_{12} = \epsilon_{21} = \gamma$, chemical	potentials $\mu_L = 5\gamma$ and $\mu_R = -5\gamma$	corresponding to a voltage bias $\Delta\mu = 10\gamma$,	and equal reservoir temperatures $T_L = T_R = 0.1 \gamma$.
	
	For the initial states shown in	Figs.~\ref{fig1}(a), (c), and (d), a consistent picture	emerges across both regimes. In the WBL, both	$\mathcal{D}(\rho_{AB})$ and $\mathcal{C}(\rho_{AB})$	evolve smoothly and saturate at a common steady-state	value of $\sim 0.14$, with $\mathcal{D}(\rho_{AB})$ and $\mathcal{C}(\rho_{AB})$ remaining comparable throughout	the evolution. In the non-Markovian regime, both	quantities exhibit large-amplitude oscillations, but their subsequent evolution differs	substantially as $\mathcal{C}(\rho_{AB})$ undergoes faster decay compared to its corresponding $\mathcal{D}(\rho_{AB})$. The $|01\rangle\langle01|$ and $|10\rangle\langle10|$ states [Fig.~\ref{fig1}(a) and (d)] build up both correlations from zero	through reservoir-generated dynamics, with non-Markovian	peaks reaching $\sim 0.5$. The  $\frac{1}{\sqrt{2}}(|10\rangle + |01\rangle)	\frac{1}{\sqrt{2}}(\langle10| + \langle01|)$ state	[Fig.~\ref{fig1}(c)] exhibits a large initial value of	$\sim 0.65$ for both $\mathcal{D}(\rho_{AB})$ and	$\mathcal{C}(\rho_{AB})$, reflecting the pre-existing	correlations encoded in the initial state prior to any	reservoir interaction. In the WBL, $\mathcal{D}(\rho_{AB})$ rapidly dissipates from its initial correlation, while the non-Markovian dynamics
	sustain large oscillations. Across all three panels, $\mathcal{D}(\rho_{AB})$ and	$\mathcal{C}(\rho_{AB})$ in the WBL overlap closely	throughout the evolution, confirming that Markovian	transport generates quantum and classical correlations	in roughly equal measure for all the considered states.
	
	Fig.~\ref{fig1}(b) illustrates $\mathcal{D}(\rho_{AB})$	and $\mathcal{C}(\rho_{AB})$ for the classical mixture
	$\rho_{23}^{\text{mix}} (0) = \frac{1}{2}(|10\rangle\langle10| +	|01\rangle\langle01|)$ as initial state, which shows a qualitatively	different behavior from all other initial states. In the	WBL, $\mathcal{C}(\rho_{AB})$ starts from $\sim 0.7$ and	decays rapidly, while	$\mathcal{D}(\rho_{AB})$ starts from zero and rises to a steady-state value,
	with both quantities converging to comparable values at	long times. In the non-Markovian regime, however,
	$\mathcal{D}(\rho_{AB})$ remains essentially zero	throughout the entire evolution with a steady-state value
	of $\sim 0.002$, while $\mathcal{C}(\rho_{AB})$ also	starts from $\sim 0.7$ but decays more slowly before
	vanishing at long times. 	The physical origin of this striking contrast between the WBL and non-Markovian behavior lies in the absence of initial coherence and the equal populations in	both single-occupation configurations $|10\rangle$ and	$|01\rangle$. When the system evolves under interdot	tunneling and reservoir interactions, each initial	component of $\rho_{23}^{\text{mix}} (t)$ independently generates coherence:	$\rho_{23}^{\text{mix}}(t) = \frac{1}{2}[\rho_{23}^{|10\rangle}(t)+\rho_{23}^{|01\rangle}(t)]$ (Appendix~\ref{app:discord}). In the isolated limit,	$\rho_{23}(t)$ from $|10\rangle$ and $|01\rangle$	oscillate with opposite phases, leading to exact cancellation, i.e., $|\rho_{23}^{\text{mix}}(t)|^2 = 0$, for the mixed state. In the non-Markovian (narrow-bandwidth) regime, the	opposite-phase relationship is preserved over extended	timescales with negligible contribution from the noise correlation, thereby maintaining near-cancellation and keeping discord close to zero.
	Since quantum discord depends more sensitively on the	$\rho_{23}$ than classical correlation, which
	retains contributions from population-based correlations,	discord is suppressed more in this regime. In the WBL, reservoir memory	is absent and the cancellation between the opposite phases is overshadowed by the contribution from the noise correlation, driving the discord towards a steady-state value of $\sim 0.14$. Both $\mathcal{D}(\rho_{AB})$ and	$\mathcal{C}(\rho_{AB})$ recover to comparable finite	values, confirming that transport-driven dynamics restore the quantum correlations that are absent in the non-Markovian case for the mixed initial state.
	
	The inset of	Fig.~\ref{fig1}(b) visualizes	steady-state universality by overlaying	$\mathcal{D}(\rho_{AB})$ for all four initial states. Despite dramatically different transient behaviors, all curves converge to the same
	steady-state value at long times, confirming that the nonequilibrium steady state is determined entirely by
	transport and reservoir parameters, independent of the initial state.

	\begin{figure*}[hbt!]
		\centering
		\begin{subfigure}[t]{0.495\textwidth}
			\centering
			\includegraphics[width=\linewidth]{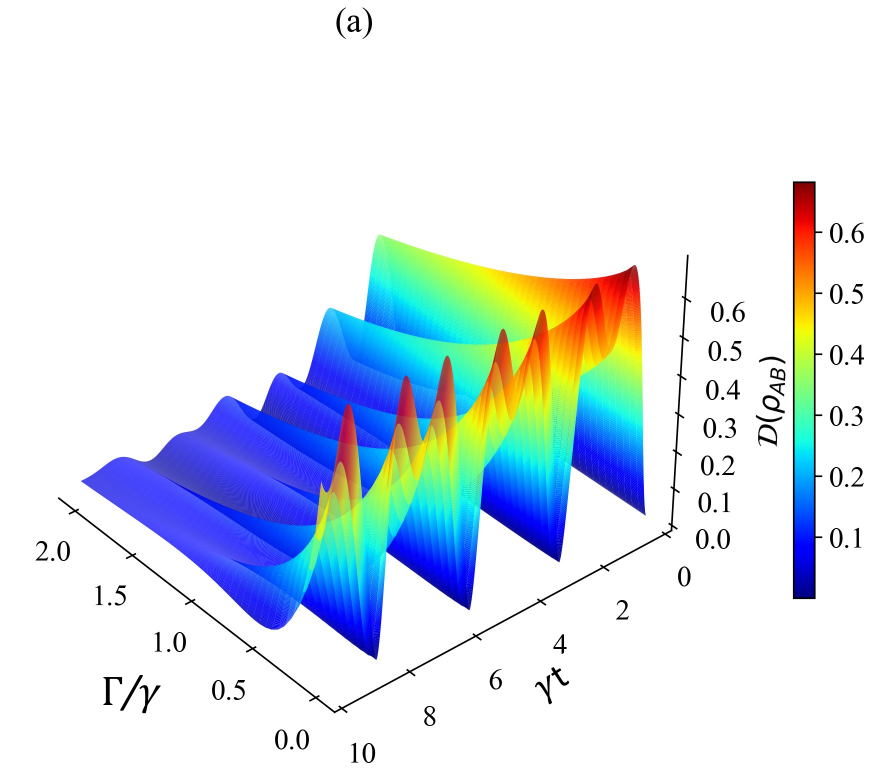}
			\label{fig:subfig1}
		\end{subfigure}
		\begin{subfigure}[t]{0.495\textwidth}
			\centering
			\includegraphics[width=\linewidth]{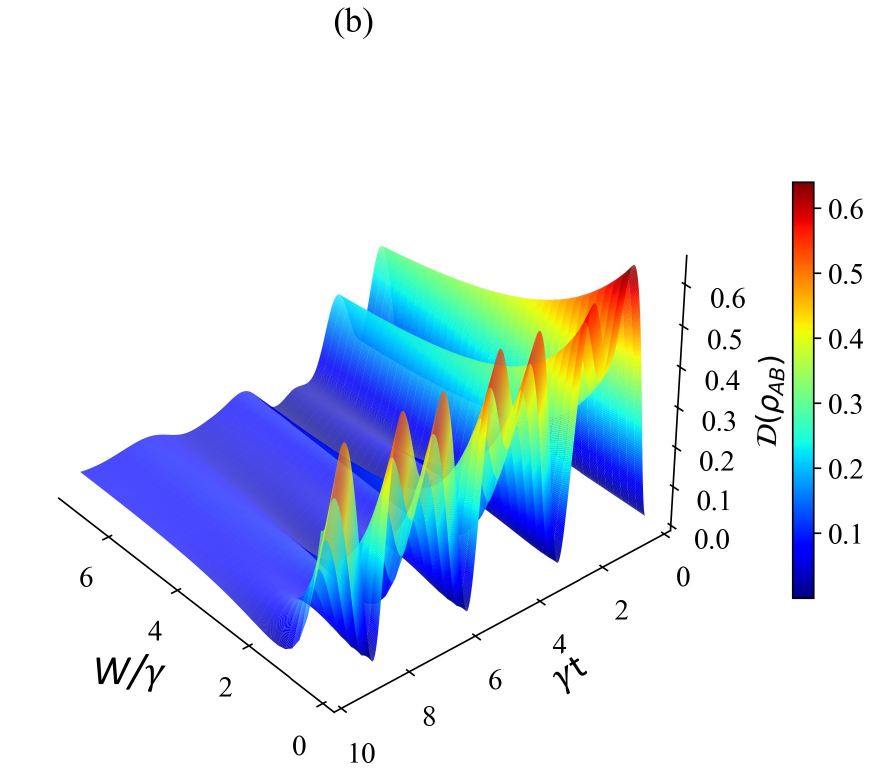}
			\label{fig:subfig2}
		\end{subfigure}
		\caption{Time evolution of quantum discord $\mathcal{D}(\rho_{AB})$ in a tunnel-coupled DQD system connected to fermionic reservoirs: (a) variation with system–reservoir coupling strength $\Gamma_L = \Gamma_R$ for fixed spectral bandwidth $W_L = W_R = W = \gamma$, and (b) variation with spectral bandwidth $W_L = W_R$ for fixed coupling strength $\Gamma_L = \Gamma_R = \Gamma = 0.5\gamma$.}
		\label{fig2}
	\end{figure*}

	In Fig.~\ref{fig2}, we illustrate the time evolution of quantum discord $\mathcal{D}(\rho_{AB})$ as a function of system-reservoir coupling strength and spectral bandwidth. The initial state of the system is taken as the single-occupation state $|01\rangle$, which means the second dot is occupied and the first dot is empty. Fig.~\ref{fig2}(a) shows the time evolution of $\mathcal{D}(\rho_{AB})$ as a function of the coupling strength $\Gamma_L = \Gamma_R \equiv \Gamma \in [0, 2.5\gamma]$ at a fixed spectral bandwidth $W_L = W_R = W = \gamma$. 
	$\mathcal{D}(\rho_{AB})$ exhibits pronounced oscillations with large amplitude and slow decay in the weak-coupling regime. 
	As $\Gamma$ increases, dissipation increasingly suppresses oscillatory dynamics which drive the system towards its steady state faster. In the strong-coupling regime, the discord goes through rapid damping to a finite steady-state due to the increasing reservoir-induced dissipation.

\begin{figure}[hbt!]
		\centering
		\includegraphics[width=0.9\textwidth]{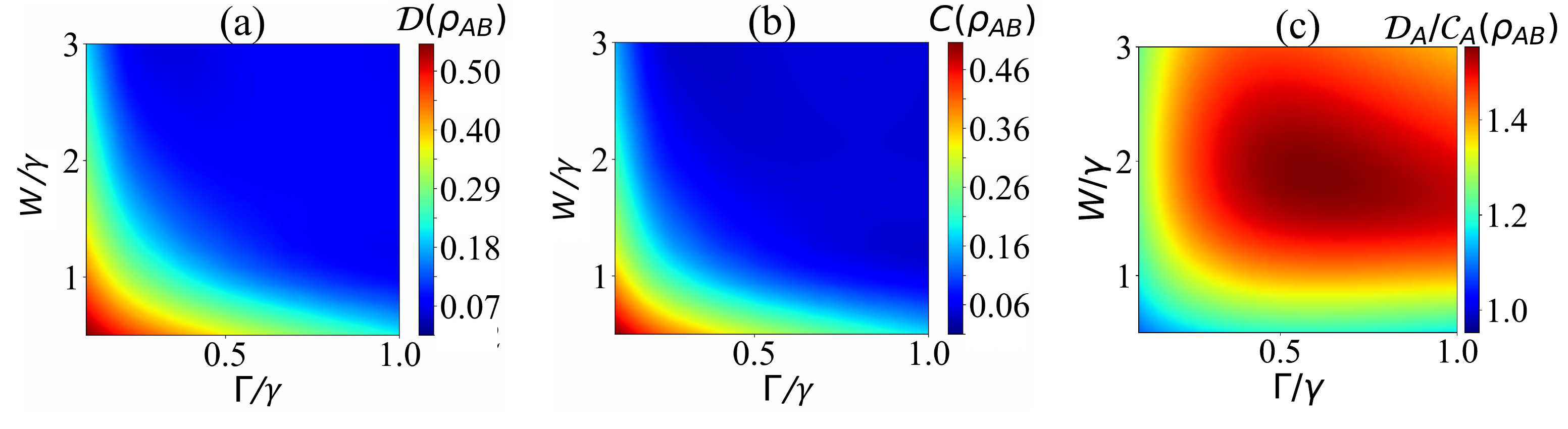}
		\caption{Steady-state: (a) quantum discord $\mathcal{D}(\rho_{AB})$, (b) classical correlation $C(\rho_{AB})$ and (c) Steady-state ratio $\mathcal{D}(\rho_{AB})/\mathcal{C}(\rho_{AB})$ as functions of coupling strength $\Gamma$ and spectral bandwidth $W$ for symmetric coupling $\Gamma_L = \Gamma_R \equiv \Gamma$ and $W_L = W_R \equiv W$.}
		\label{fig3}
	\end{figure}

\begin{figure}[hbt!]
		\centering
		\includegraphics[width=0.7\textwidth]{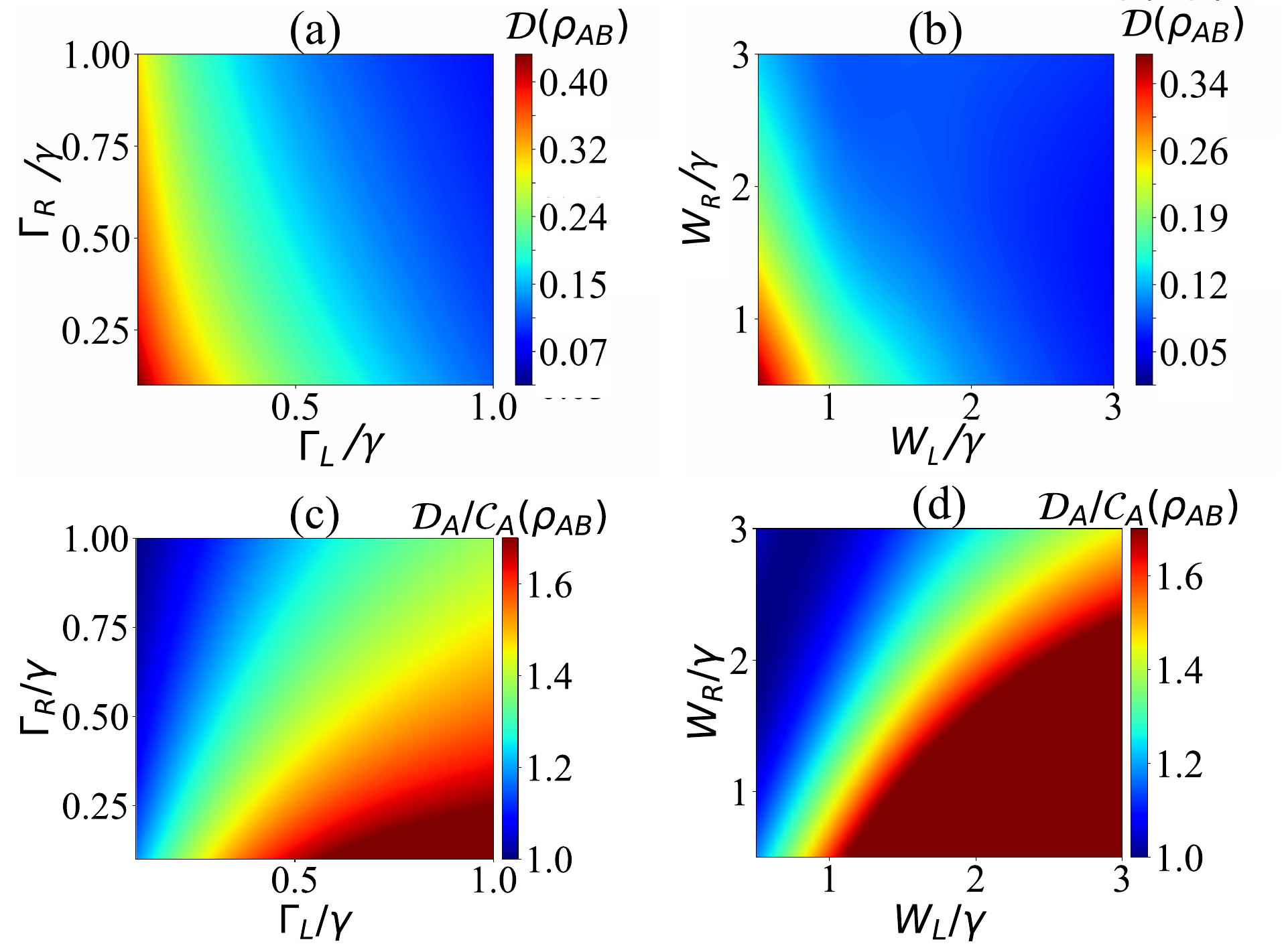}
		\caption{Steady-state quantum discord $\mathcal{D}(\rho_{AB})$ as a function of 
			(a) left and right coupling strengths $(\Gamma_L, \Gamma_R)$ at fixed spectral 
			bandwidth $W_L = W_R = \gamma$, and (b) left and right spectral bandwidths 
			$(W_L, W_R)$ at fixed coupling strength $\Gamma_L = \Gamma_R = 0.5\gamma$. The corresponding 
			steady-state ratio $\mathcal{D}_A/\mathcal{C}_A(\rho_{AB})$ as functions of 
			(c) $(\Gamma_L, \Gamma_R)$ and (d) $(W_L, W_R)$ at the same fixed parameters 
			as (a) and (b) respectively, for symmetric initial state $|01\rangle\langle 01|$.}
		\label{fig4}
	\end{figure}

	Fig.~\ref{fig2}(b) shows the time evolution of $\mathcal{D}(\rho_{AB})$ as a function of spectral bandwidth $W_L = W_R \equiv W$ at fixed coupling strength $\Gamma_L = \Gamma_R = \Gamma = 0.5\gamma$. In the narrow-bandwidth regime, the sharply peaked Lorentzian spectral density restricts the available energy levels for particle/energy exchange to a
	narrow window of reservoir frequencies, which strongly amplifies non-Markovian memory effects. This results in large-amplitude, high-frequency oscillations with pronounced partial revivals, which are qualitatively similar to those observed in the weak-coupling regime of Fig.~\ref{fig2}(a). 
	As $W$ increases, the spectral density broadens and flattens, eventually decreasing memory effects and driving the system dynamics towards effectively Markovian behavior. In the large-bandwidth regime, $\mathcal{D}(\rho_{AB})$ relaxes rapidly towards a steady-state value without much oscillation. This behavior is similar to the effect of increasing coupling strength.
	However, the two parameters affect the system differently. Increasing $\Gamma$ primarily enhances dissipation by increasing the strength of system-reservoir interactions, while increasing $W$ primarily destroys reservoir memory due to the increase in reservoir energy levels interacting with the system. Despite these differences, quantum discord remains finite in the steady state across the entire bandwidth range, further reinforcing its robustness as a non-classical resource induced by the nonequilibrium transport in the open system. 
	The role of spectral bandwidth and coupling strength in jointly controlling steady-state correlations is systematically examined in Figs.~\ref{fig3} and~\ref{fig4}.

	Fig.~\ref{fig3} shows the steady-state quantum discord $\mathcal{D}(\rho_{AB})$, classical correlation
	$\mathcal{C}(\rho_{AB})$, and their ratio $\mathcal{D}(\rho_{AB})/\mathcal{C}(\rho_{AB})$ as a function of the symmetric coupling strength	$\Gamma_L = \Gamma_R \equiv \Gamma$ and spectral bandwidth $W_L = W_R \equiv W$, with	$|01\rangle\langle01|$ as initial state. These three plots present a comprehensive map of how nonclassical and classical correlations, and their relative magnitudes, respond to environmental parameters
	in the nonequilibrium steady state.
	
	In Fig.~\ref{fig3}(a), the steady-state quantum	discord $\mathcal{D}(\rho_{AB})$ is shown as a function of
	$\Gamma$ and $W$. In the weak-coupling, narrow-bandwidth regime ($\Gamma/\gamma \ll 1$, $W/\gamma \ll 1$), $\mathcal{D}(\rho_{AB})$ reaches its maximum value of $\sim 0.5$, where non-Markovian memory effects are strongest and reservoir-induced	dissipation is minimal. $\mathcal{D}(\rho_{AB})$  decreases smoothly as either $\Gamma$ or $W$	increases, reflecting the progressive suppression of quantum correlations by enhanced dissipation.
	Importantly, $\mathcal{D}(\rho_{AB})$ sustains a nonzero value everywhere in the parameter space shown, reaching a finite saturation value of $\sim 0.07$. This finite value signifies the importance of nonequilibrium transport in preserving quantum correlations, as long as the system is subjected to a finite chemical bias and particle current flows through the double dot.
	
	Fig.~\ref{fig3}(b) shows the corresponding	steady-state classical correlation	$\mathcal{C}(\rho_{AB})$. While the overall	structure mirrors that of Fig.~\ref{fig3}(a), with the	maximum value of $\sim 0.46$ concentrated in
	the weak-coupling, narrow-bandwidth corner and	decreasing with increasing $\Gamma$ and $W$, classical correlations decay significantly faster than quantum discord across the	parameter space. This differential decay reflects	the greater sensitivity of classically accessible	correlations to environmental noise and	dissipation.
	
	Fig.~\ref{fig3}(c) shows the ratio	$\mathcal{D}(\rho_{AB})/\mathcal{C}(\rho_{AB})$	across the same parameter space. As far as the $|01\rangle$ initial state is concerned, the ratio exceeds unity everywhere, confirming that	quantum discord always equals or exceeds	classical correlation regardless of coupling strength	or bandwidth. In the weak-coupling,	narrow-bandwidth corner, the ratio approaches	$\sim 1.0$, indicating that both types of	correlations are of comparable magnitude. As $\Gamma$ and $W$ increase,
	the ratio rises monotonically toward $\sim 1.4$--$1.5$, reflecting the faster decay of classical correlations relative to discord established in Fig.~\ref{fig3}(a) and (b), and upon further increase in both parameters, the ratio begins to decrease. This monotonic behavior demonstrates that increasing either the coupling strength or the bandwidth progressively enhances the relative quantum character of the steady-state correlations, even as their absolute magnitudes decrease. The three plots collectively establish a clear operational framework: the weak-coupling, narrow-bandwidth regime maximizes the magnitude of both discord and classical correlations, while the coupling in the range $\sim (0.4 - 0.9)\gamma$ and bandwidth in the range $\sim (1.5 - 2.5)\gamma$ maximizes their relative quantum dominance.
	
	For the initial state $|01\rangle\langle01|$, the influence of reservoir asymmetry on steady-state quantum discord $\mathcal{D}(\rho_{AB})$ and the ratio	$\mathcal{D}(\rho_{AB})/\mathcal{C}(\rho_{AB})$ is explored in Fig.~\ref{fig4}. In Fig.~\ref{fig4}(a), $\mathcal{D}(\rho_{AB})$ is shown as a function of $\Gamma_L$ and $\Gamma_R$ at fixed $W_L = W_R = \gamma$, while Fig.~\ref{fig4}(b) shows $\mathcal{D}(\rho_{AB})$ as a function of $W_L$ and $W_R$ at fixed $\Gamma_L = \Gamma_R = 0.5\gamma$. In Fig.~\ref{fig4}(a), the discord contours are nearly vertical across the entire parameter space, indicating that $\mathcal{D}(\rho_{AB})$ is governed almost entirely by $\Gamma_L$ and is largely insensitive to $\Gamma_R$. The discord attains its maximum value of $\sim 0.36$ at small $\Gamma_L$ and decreases rapidly as $\Gamma_L$ increases, while varying $\Gamma_R$ across its full range produces negligible change. Fig.~\ref{fig4}(b) reveals the same pronounced left-right asymmetry in the bandwidth dependence, with $\mathcal{D}(\rho_{AB})$ controlled predominantly by $W_L$ and $W_R$ playing a negligible role. This asymmetry can be attributed to the proximity of the dot energy levels to those of the left reservoir: since $\epsilon_{11} = 3\gamma$ and $\epsilon_{22} = 2\gamma$ lie closer to $\mu_L = 5\gamma$ than to $\mu_R = -5\gamma$, the left reservoir couples more closely to the dots, making it the dominant channel for particle exchange.
	
	Fig.~\ref{fig4}(c) and~\ref{fig4}(d) show the	corresponding steady-state ratio	$\mathcal{D}(\rho_{AB})/\mathcal{C}(\rho_{AB})$	as functions of $(\Gamma_L, \Gamma_R)$ and	$(W_L, W_R)$ at the same fixed parameters as	Fig.~\ref{fig4} (a) and (b) respectively. In contrast to the nearly vertical contours of	Fig.~\ref{fig4} (a) and (b), the ratio contours acquire	a pronounced diagonal orientation in both panels,	revealing that the relative dominance of quantum over classical correlations is controlled by	both reservoirs simultaneously with opposing roles. In Fig.~\ref{fig4}(c), the ratio decreases from	$\sim 1.6$ in the small-$\Gamma_L$,	large-$\Gamma_R$ corner to $\sim 1.0$ in the	large-$\Gamma_L$, small-$\Gamma_R$ corner.	Increasing $\Gamma_R$ raises the ratio while	increasing $\Gamma_L$ lowers it, which is directly	opposite to their roles in controlling absolute	discord in Fig.~\ref{fig4}(a). 	Fig.~\ref{fig4}(d) shows qualitatively	similar but distinct behavior for the bandwidth	dependence. The ratio is largest ($\sim 1.6$)	in the large-$W_L$, small-$W_R$ corner and	smallest ($\sim 1.0$) in the small-$W_L$,	large-$W_R$ corner, with $W_L$ and $W_R$	again playing opposing roles. As in the coupling	case, $W_R$, which is less sensitive for absolute	discord in Fig.~\ref{fig4}(b), emerges as an equally important parameter for quantum dominance.
	
	Together, Fig.~\ref{fig4}(c) and (d) reveal a rich	complementarity between the two reservoirs:	the left reservoir governs the absolute	magnitude of discord, while both reservoirs	jointly and opposingly determine its dominance	over classical correlations, providing	additional experimental degrees of freedom	for independently engineering the magnitude	and quantum character of steady-state	correlations.

	\begin{figure}[h]
		\centering
		\includegraphics[width=0.95\textwidth]{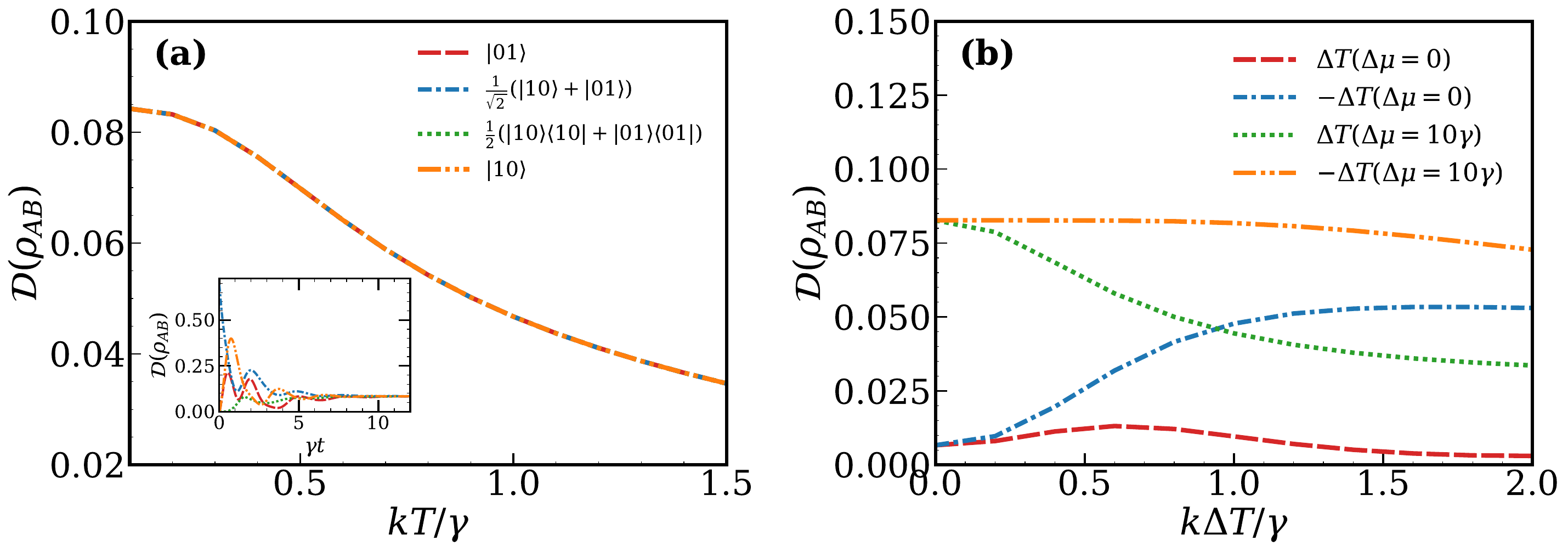}
		\caption{(a) For both reservoirs at the same temperatures $T_L = T_R = T$, the steady-state quantum discord $\mathcal{D}(\rho_{AB})$ as a function of
			temperature $ k T/\gamma$ for four initial states
			$|01\rangle$,
			$\frac{1}{\sqrt{2}}(|10\rangle + |01\rangle)$,
			$\frac{1}{2}(|10\rangle\langle10| + |01\rangle\langle01|)$,
			and $|10\rangle$, (b) the steady-state
			$\mathcal{D}(\rho_{AB})$ as a function of thermal
			bias $\Delta T = T_L - T_R$ for voltage biases
			$\Delta\mu = 0$ and $\Delta\mu = 10\gamma$, and
			for both positive ($\Delta T > 0$, left reservoir
			hotter) and negative ($\Delta T < 0$, right
			reservoir hotter) thermal gradients.}
		\label{fig5}
	\end{figure}

	Fig.~\ref{fig5} examines the influence of temperature	on steady-state quantum discord and its dependence on	thermal bias and voltage bias, with	$|01\rangle\langle01|$ as the initial state. In Fig.~\ref{fig5}(a), the steady-state	$\mathcal{D}(\rho_{AB})$ is shown as a function of temperature	$kT/\gamma$ for the four initial states	$|01\rangle\langle01|$,	$\frac{1}{\sqrt{2}}(|10\rangle+|01\rangle)	\frac{1}{\sqrt{2}}(\langle10|+\langle01|)$,	$\frac{1}{2}(|10\rangle\langle10|+|01\rangle\langle01|)$,	and $|10\rangle\langle10|$. The four curves overlap perfectly across the entire temperature	range, confirming again that the	nonequilibrium steady-state discord is completely	independent of the initial state. The steady-state	discord decreases monotonically with increasing	temperature, indicating the progressive suppression	of quantum correlations by thermally enhanced	dissipation.	The inset shows the time evolution of	$\mathcal{D}(\rho_{AB})$ for all four initial	states, directly illustrating the origin of this	overlap where, despite the different transient	behaviors, all curves converge to the same	steady-state value as the initial-state dependence of $\rho_{23}(t)$ decays and	the long-time dynamics is governed entirely	by the noise-driven term $v_{12}$, which depends	only on reservoir parameters and nonequilibrium	driving.
	
	Fig.~\ref{fig5}(b) shows the steady-state	$\mathcal{D}(\rho_{AB})$ as a function of thermal	bias $\Delta T = T_L - T_R$ for voltage biases	$\Delta\mu = 0$ and $\Delta\mu = 10\gamma$, and	for both positive ($\Delta T > 0$, left reservoir	hotter) and negative ($\Delta T < 0$, right	reservoir hotter) thermal gradients. At	$\Delta\mu = 10\gamma$, both cases start from	$\sim 0.084$ at $\Delta T = 0$. For positive	$\Delta T$ (green dotted), discord decreases	monotonically and significantly, reflecting the	strong decoherence effect of the hotter left	reservoir. For negative	$\Delta T$ (orange dash-dot-dot), discord decreases	slowly and remains comparatively robust across the	entire range of $|\Delta T|$. This asymmetry is consistent with the left-reservoir dominance established in  Fig.~\ref{fig4}.

	At $\Delta\mu = 0$, for positive $\Delta T$	(red dashed), discord remains near zero	($\lesssim 0.015$) across the entire range. For negative	$\Delta T$ (blue dash-dot), discord increases	monotonically to $\sim 0.052$ at	$|k\Delta T| = 2\gamma$, demonstrating that a	thermal gradient with the right reservoir hotter
	can generate finite quantum discord through	thermally driven transport even in the complete	absence of voltage bias.

	\section{Conclusions}
	
	The central finding of this work is that nonequilibrium transport, rather than initial-state
	preparation, determines the quantum correlations that survive in the long-time steady state. Across
	all four initial states considered, the steady-state discord converges to a common value set entirely by reservoir parameters and the applied voltage bias. 
	For the classical mixture initial state, the discord in the non-Markovian (narrow-bandwidth) regime remains
	suppressed near zero throughout the transient, while the other initial states build up substantial
	correlations. The mechanism is a phase cancellation: the $\ket{10}$ and $\ket{01}$ components
	each independently generate coherence $\rho_{23}(t)$, but with opposite phases, so their
	contributions to the mixture nearly cancel. The narrow-bandwidth reservoir preserves this
	cancellation over extended timescales, keeping discord close to zero. In the wideband limit,
	the same cancellation is overwhelmed by noise-correlation contributions that drive the system
	toward the universal steady-state value of $\sim 0.14$, the same value reached by every other
	initial state.
	
	The transition from non-Markovian to Markovian behavior, driven by increasing coupling strength
	$\Gamma$ or spectral bandwidth $W$, tells a consistent story about how the environment shapes
	quantum correlations. Both parameters suppress the oscillatory transient and accelerate
	relaxation, but through distinct channels: $\Gamma$ increases the overall dissipation rate, while
	$W$ broadens the Lorentzian spectral density. Despite this
	suppression, discord never vanishes in the steady state across the parameter space explored,
	and the ratio $\mathcal{D}/\mathcal{C}$ grows with both $\Gamma$ and $W$, reaching $\sim 1.5$
	at the largest values considered. Classical correlations decay faster than quantum discord under
	increasing dissipation, so the environment enhances the relative quantum
	character of the steady state even as it reduces the absolute magnitudes.
	
	The left-right reservoir asymmetry adds a further layer of structure. Because the dot energy
	levels lie closer to the left chemical potential, the left reservoir has more influence on the absolute
	magnitude of discord, while the right reservoir
	turns out to play a prominent role for tuning the ratio $\mathcal{D}/\mathcal{C}$. The two
	reservoirs thus play complementary roles, offering separate experimental
	knobs for controlling the magnitude and quantum character of correlations without one
	compromising the other.
	
	Finally, thermal bias reveals that the direction of the heat current matters as much as its
	magnitude. At finite voltage bias, heating the left reservoir suppresses discord strongly while
	heating the right has a comparatively mild effect, again reflecting left-reservoir dominance.
	More strikingly, at zero voltage bias, a thermal gradient can generate finite discord on its
	own, but only when the right reservoir is hotter: the reverse configuration keeps discord
	negligibly small. Together, these findings suggest that spectral engineering, coupling
	asymmetry, and thermal-gradient control constitute a practical toolkit for controlling quantum
	correlations in open fermionic devices.
	
	\section{Acknowledgments}
	
	The authors (TYM, SK, and AS) gratefully acknowledge the financial support of the Anusandhan National Research Foundation (ANRF), Government of India [Project File No. CRG/2022/007836]. All the computational resources provided by the ANRF-CRG funded GPGPU system along with the SRM-HPCC at SRMIST, India, are also gratefully acknowledged.
	
	\appendix
	\section*{Appendix: Analytical Evaluation of classical correlation for the Double Quantum Dot}
	\label{app:discord}
	
	We present the explicit calculation of quantum discord for the restricted X-state density matrix of the double quantum dot system, as outlined in Eq.~(19), following the procedure for spinless fermionic two-dot systems~\cite{krithivasan2025interplay}.
	
	\subsection*{A.1 \quad Density Matrix and Reduced States}
	
	In the occupation-number basis $\{|11\rangle, |10\rangle, |01\rangle, |00\rangle\}$, the normalized density matrix is
	\begin{equation}
		\rho(t) =
		\begin{pmatrix}
			a & 0 & 0 & 0 \\
			0 & b & z & 0 \\
			0 & z^* & c & 0 \\
			0 & 0 & 0 & d
		\end{pmatrix},
	\end{equation}
	where $a = \rho_{11}$, $b = \rho_{22}$, $c = \rho_{33}$, $d = \rho_{44}$, $z = \rho_{23}$, and the normalization condition is $a+b+c+d = 1$.
	The reduced density matrices obtained by partial trace are
	\begin{align}
		\rho_A = \mathrm{Tr}_B(\rho) &=
		\begin{pmatrix} a+b & 0 \\ 0 & c+d \end{pmatrix}, \\
		\rho_B = \mathrm{Tr}_A(\rho) &=
		\begin{pmatrix} a+c & 0 \\ 0 & b+d \end{pmatrix}.
	\end{align}
	Defining the binary entropy $h(x) = -x\log_2 x - (1-x)\log_2(1-x)$, the von Neumann entropies are
	\begin{align}
		S(\rho_A) &= h(a+b), \\
		S(\rho_B) &= h(a+c).
	\end{align}
	
	\subsection*{A.2 \quad Total Mutual Information}
	
	The eigenvalues of $\rho$ are $\lambda_3 = a$ and $\lambda_4 = d$ from the two $1\times1$ blocks, and
	\begin{equation}
		\lambda_{1,2} = \frac{(b+c) \pm \sqrt{(b-c)^2 + 4|z|^2}}{2},
	\end{equation}
	from the $2\times2$ block. The von Neumann entropy is
	\begin{equation}
		S(\rho) = -\sum_{i=1}^{4} \lambda_i \log_2 \lambda_i,
	\end{equation}
	and the total mutual information is
	\begin{equation}
		\mathcal{I}(\rho_{AB}) = S(\rho_A) + S(\rho_B) - S(\rho).
	\end{equation}
	
	\subsection*{A.3 \quad Classical Correlations via Optimized Measurements}
	
	As per Eq.~(\ref{eq:classical}) in the main text, the classical correlation is defined by performing projective
	measurements on subsystem $A$ (first dot). Consequently, the post-measurement states are states
	of $B$ (second dot). The relevant single-dot basis for dot $A$ is the occupation basis 
	$\{|1\rangle,|0\rangle\}$, where $|1\rangle$ denotes an occupied dot and $|0\rangle$ an empty dot.
	A general rank-one projective measurement on $A$ is characterised by Bloch sphere angles $(\theta, \phi)$
	and is expressed in this basis as
	
	\begin{equation}
		|\psi_+\rangle = \cos\frac{\theta}{2}|1\rangle + e^{i\phi}\sin\frac{\theta}{2}|0\rangle,
	\end{equation}
	
	\begin{equation}
		|\psi_-\rangle = \sin\frac{\theta}{2}|1\rangle - e^{i\phi}\cos\frac{\theta}{2}|0\rangle,
	\end{equation}
	
	with $\Pi^A_k = |\psi_k\rangle\langle\psi_k|$, satisfying $\Pi^A_+ + \Pi^A_- = \mathbb{I}_A$ 
	and $\Pi^A_+\Pi^A_- = 0$. Their explicit matrix forms in the $\{|1\rangle,|0\rangle\}$ basis are
	
	\begin{equation}
		\Pi^A_+ =
		\begin{pmatrix}
			\cos^2\dfrac{\theta}{2} & e^{-i\phi}\cos\dfrac{\theta}{2}\sin\dfrac{\theta}{2} \\[8pt]
			e^{i\phi}\cos\dfrac{\theta}{2}\sin\dfrac{\theta}{2} & \sin^2\dfrac{\theta}{2}
		\end{pmatrix},
	\end{equation}
	
	\begin{equation}
		\Pi^A_- =
		\begin{pmatrix}
			\sin^2\dfrac{\theta}{2} & -e^{-i\phi}\sin\dfrac{\theta}{2}\cos\dfrac{\theta}{2} \\[8pt]
			-e^{i\phi}\sin\dfrac{\theta}{2}\cos\dfrac{\theta}{2} & \cos^2\dfrac{\theta}{2}
		\end{pmatrix}.
	\end{equation}
	
	For each outcome $k \in \{+,-\}$, the probability and post-measurement state of $B$ are given by:
	
	\begin{equation}
		p_k = \mathrm{Tr}\!\left[
		\left(\Pi^A_k \otimes \mathbb{I}_B\right)\rho_{AB}
		\left(\Pi^A_k \otimes \mathbb{I}_B\right)
		\right],
	\end{equation}
	
	\begin{equation}
		\rho^k_B = \frac{
			\mathrm{Tr}_A\!\left[
			\left(\Pi^A_k \otimes \mathbb{I}_B\right)\rho_{AB}
			\left(\Pi^A_k \otimes \mathbb{I}_B\right)
			\right]
		}{p_k}.
	\end{equation}
	
	The conditional entropy is $S(\rho_{B|A}) = \sum_k p_k\, S(\rho^k_B)$. For the X-state structure
	of $\rho$, the phase $\phi$ drops out after minimization, reducing the optimization to a single
	parameter $\theta$. Since the measurement is on $A$, whose reduced state is
	$\rho_A = \mathrm{diag}(a+b,\, c+d)$, the outcome probabilities are
	
	\begin{equation}
		p_+ = (a+b)\cos^2\frac{\theta}{2} + (c+d)\sin^2\frac{\theta}{2},
	\end{equation}
	
	\begin{equation}
		p_- = (a+b)\sin^2\frac{\theta}{2} + (c+d)\cos^2\frac{\theta}{2} \;=\; 1 - p_+.
	\end{equation}
	
	Each post-measurement state $\rho^k_B$ is a $2\times 2$ diagonal matrix, and the conditional
	entropy takes the form
	
	\begin{equation}
		S(\rho_{B|A},\,\theta) = p_+\, h\!\left(\frac{1+\xi_+}{2}\right)
		+ p_-\, h\!\left(\frac{1+\xi_-}{2}\right),
	\end{equation}
	
	where
	
	\begin{equation}
		\xi_{\pm} = \frac{
			(a - d)\cos\theta \;\pm\;
			\sqrt{(c - b)^2\cos^2\theta + 4|z|^2\sin^2\theta}
		}{2p_{\pm}}.
	\end{equation}
	
	The classical correlation is then
	
	\begin{equation}
		\mathcal{C}(\rho_{AB}) = S(\rho_B) - \min_{\theta\in[0,\,\pi/2]} S(\rho_{B|A},\,\theta),
	\end{equation}
	
	consistent with Eq.~(\ref{eq:classical}). 
	
	\subsection*{A.4  \quad  $\rho_{23}^{\text{mix}}(t) $}
	
	We consider a maximally mixed, single-occupation initial state, defined as
	\begin{equation}
		\rho^{\text{mix}}(0) = \frac{1}{2}\left(|10\rangle\langle 10| + |01\rangle\langle 01|\right).
	\end{equation}
	The corresponding coherence $\rho_{23}(t) = \langle a_2^\dagger(t) a_1(t)\rangle$ satisfies
	\begin{equation}
		\rho_{23}^{\text{mix}}(t) = \frac{1}{2}\left[\rho_{23}^{|10\rangle}(t) + \rho_{23}^{|01\rangle}(t)\right] 
		\label{eq:A1}
	\end{equation}
	for all times $t$, all coupling strengths, and all bath parameters. The same averaging holds for any observable linear in the initial correlation matrix.
	
	The Heisenberg equations are linear because the total Hamiltonian is quadratic. Hence the quantum Langevin equation, Eq.~\ref{eq:langevin}, admits the exact solution
	\begin{equation}
		a_i(t) = \sum_j u_{ij}(t,t_0) a_j(t_0) + F_i(t), 
		\label{eq:A2}
	\end{equation}
	where $u_{ij}$ is a Green's function independent of the initial system state, and $F_i(t)$ depends only on bath operators. The initial state is factorized as $\rho_{\text{total}}(0) = \rho_S(0) \otimes \rho_B$ with $\rho_B$ thermal, so $\langle F_i\rangle = 0$.
	
	Expanding $\rho_{23}(t) = \langle a_2^\dagger a_1\rangle$ using Eq.~(\ref{eq:A2}) gives four terms. Taking the expectation value:
	\begin{itemize}
		\item $\langle F_i\rangle = 0$ eliminates terms linear in $F_i$.
		\item Cross terms factorize as $\langle a_m^\dagger\rangle\langle F_1\rangle = 0$. 
		\item $\langle F_2^\dagger F_1\rangle$ is independent of $\rho_S(0)$.
	\end{itemize}
	Only the first sum survives:
	\begin{equation}
		\rho_{23}(t) = \sum_{m,n} u_{2m}^* u_{1n} \langle a_m^\dagger a_n\rangle_0 + B(t),
		\label{eq:A3}
	\end{equation}
	where $B(t)=\langle F_2^\dagger F_1\rangle$ . Eq.~(\ref{eq:A3}) is linear in the initial correlation matrix $\langle a_m^\dagger a_n\rangle_0$.
	
	For $|10\rangle$, $|01\rangle$, and $\rho_{\text{mix}}$, the only nonzero initial correlations are the populations:
	\begin{equation}
		\begin{array}{c|cc}
			& \langle a_1^\dagger a_1\rangle_0 & \langle a_2^\dagger a_2\rangle_0 \\ \hline
			|10\rangle & 1 & 0 \\
			|01\rangle & 0 & 1 \\
			\rho^{\text{mix}}_{(0)} & 1/2 & 1/2
		\end{array}
	\end{equation}
	All off-diagonal correlations are zero. Substituting into Eq.~(\ref{eq:A3}) yields
	\begin{align}
		\rho_{23}^{|10 \rangle}(t) &= u_{21}^*(t)u_{11}(t) + B(t), \\
		\rho_{23}^{|01\rangle}(t) &= u_{22}^*(t)u_{12}(t) + B(t), \\
		\rho_{23}^{\text{mix}}(t) &= \tfrac12(u_{21}^*(t)u_{11}(t) + u_{22}^*(t)u_{12}(t)) + B(t),
	\end{align}
	from which Eq.~(\ref{eq:A1}) follows directly.
	
	For quartic observables such as $\rho_{11}(t) = \langle N_1 N_2\rangle$, linearity does not directly apply. However, for this specific mixture (diagonal in the number basis, no initial coherences), the averaging property continues to hold because no cross-terms between $|10\rangle$ and $|01\rangle$ contribute upon expansion. The proof is valid for all bath parameters ($\Gamma_\alpha$, $W_\alpha$, $T_\alpha$) as they enter only through $u_{ij}$ and $B(t)$, which are identical for the three initial states.
	
	\bibliographystyle{apsrev}
	
	\bibliography{Quantum_discord}
	
\end{document}